%% file: paper.tex
\documentclass[10pt,journal,compsoc]{IEEEtran}
\usepackage{balance}
\usepackage{subcaption}
\usepackage{cite}
\usepackage[usenames,dvipsnames]{color}
\usepackage{colortbl}
\definecolor{mygray}{gray}{.8}
\input{macros}

\usepackage{algorithm}
\usepackage{algorithmic}

\newcolumntype{s}{>{\hsize=.2\hsize}X}
\newcolumntype{f}{>{\hsize=.75\hsize}X}
\newcolumntype{b}{>{\hsize=.05\hsize}X}

\begin{document}

\title{Explainable Software Defect Prediction: Are We There Yet?}

\author{Jiho~Shin, Reem~Aleithan, Jaechang~Nam, Junjie~Wang, Song~Wang, Member, IEEE
\IEEEcompsocitemizethanks{
\IEEEcompsocthanksitem J. Shin is with is with Lassonde School of Engineering, York University, Canada.\protect\\
E-mail: jihoshin@yorku.ca
\IEEEcompsocthanksitem R. Aleithan is with Huawei. \protect\\
E-mail: reem.aleithan@huawei.com
\IEEEcompsocthanksitem J. Nam is with Computer Science and Electrical Engineering, Handong Global University, South Korea. \protect\\
E-mail: jcnam@handong.edu
\IEEEcompsocthanksitem J. Wang is with Institute of Software Chinese Academy of Sciences and University of Chinese Academy of Sciences, China.\protect\\
E-mail: wangjunjie@itechs.iscas.ac.cn
\IEEEcompsocthanksitem S. Wang is with Lassonde School of Engineering, York University, Canada.\protect\\
E-mail: wangsong@yorku.ca
}
\thanks{Manuscript received xxx xx, 2021; revised xxx xx, 2021.}
}

\IEEEtitleabstractindextext{%
\begin{abstract}
\input{sec/abstract}
\end{abstract}
}

\maketitle

\input{sec/introduction}

\input{sec/motivation}
\input{sec/setup}
\input{sec/result}
\input{sec/discussion}
\input{sec/related}
\input{sec/conclusion}

\bibliographystyle{IEEEtran}
\bibliography{paper}

\newpage

\end{document}

%% file: macros.tex
\usepackage{soul}
\usepackage{hyperref}
\usepackage{multirow}
\usepackage{listings}
\usepackage{fancybox}
\usepackage{enumitem}
\usepackage{graphicx}
\usepackage{color}
\usepackage{xcolor}
\usepackage{array}
\usepackage{amsmath}
\usepackage{xspace}
\usepackage{url}
\usepackage{tikz}
\usepackage{caption}
\usepackage{pgfplots}
\pgfplotsset{compat=1.16}
\usepackage{pgf-pie}
\usetikzlibrary{pgfplots.statistics,calc}
\usepackage{algorithm}
\usepackage{algorithmic}

\usepackage{tcolorbox}
\makeatletter
\newcommand{\mybox}[1]{%
	\setbox0=\hbox{#1}%
	\setlength{\@tempdima}{\dimexpr\wd0+13pt}%
	\begin{tcolorbox}[boxrule=0.5pt, colback=white, arc=4pt,
		left=6pt,right=6pt,top=6pt,bottom=6pt,boxsep=0pt]
		#1
	\end{tcolorbox}
}

\definecolor{songcolor}{RGB}{191,191,191}

\newcommand{\rd}{$rank\_di\!f\!\!f\!$}
\newcommand{\rankdiff}{$rank\_di\!f\!\!f\!$}
\newcommand{\Rankdiff}{$rank\_di\!f\!\!f\!$}
\newcommand{\hitrate}{$hit\_rate$}

%% file: sec/abstract.tex
Explaining the results of defect prediction models is practical but challenging to achieve. 
Recently, Jiarpakdee et al. \cite{jiarpakdee2020empirical} proposed to use {two state-of-the-art} model-agnostic techniques (i.e., \emph{LIME} and \emph{BreakDown}) to explain prediction results. 
Their study showed that model-agnostic techniques can achieve remarkable performance, and the generated explanations can assist developers to understand the prediction results. 
However, the fact that they only examined both \emph{LIME} and \emph{BreakDown} in a single defect prediction setting calls into question the consistency and reliability of model-agnostic techniques on defect prediction models under various settings.

In this paper, we set out to investigate the reliability and stability of explanation generation approaches based on model-agnostic techniques, i.e., \emph{LIME} and \emph{BreakDown}, on defect prediction models under different settings, e.g., data sampling techniques, machine learning classifiers, and prediction scenarios used when building defect prediction models. 
Specifically, we use both \emph{LIME} and \emph{BreakDown} to generate explanations for the same instance under various defect prediction models with different settings and then check the consistency of the generated explanations for the instance. 
{We reused the same 
defect data from Jiarpakdee et al. in our experiments.} 
The results show that both \emph{LIME} and \emph{BreakDown} generate inconsistent explanations under different defect prediction settings for the same test instances. These imply that the model-agnostic techniques are unreliable for practical explanation generation. {In addition, our manual analysis shows that none of the generated explanations can reflect the root causes of the predicted defects, which further weakens the usefulness of model-agnostic based explanation generation.}
Overall, with this study, we {urge a revisit of existing model-agnostic based studies in software engineering} and call for more research in explainable defect prediction towards achieving reliable and stable explanation generation. 

%% file: sec/introduction.tex
\section{Introduction}
\label{sec:intro}
Software Defect Prediction (SDP) models have been actively studied to allocate testing resources efficiently to reduce development costs.
Most existing SDP models use various code and development metrics as features to classify a target code fragment as buggy or not. 
However, a major issue that SDP models face is that they lack actionable messages for the developers to act upon~\cite{lewis2013does}, making it very difficult for practical usage.

To address this issue, studies investigating explainable artificial intelligence (XAI) in the domain of defect prediction have been explored recently~\cite{humphreys2019explainable, 9193975, khanan2020jitbot, JiarpakdeeTG21, pornprasit2021jitline} but most of these approaches target at global explanation, which summarizes a prediction of a whole model (i.e., the relationship between defect prediction features and the bug proneness). 
Since the global explanation does not provide a detailed interpretation of prediction results, Jiarpakdee et al.~\cite{jiarpakdee2020empirical} proposed to use the model-agnostic methods, i.e.,  \emph{LIME}~\cite{ribeiro2016should} and \emph{BreakDown}~\cite{gosiewska2019ibreakdown, staniak2018explanations} to generate instance explanation to explain the prediction of each target code fragment. 
The explanation is defined as a list of ordered features.  
Their experiments and use case studies showed that both \emph{LIME} and \emph{BreakDown} achieve promising performance and the generated explanations can assist developers by showing actionable guidance for practical usages.

However, in Jiarpakdee et al.~\cite{jiarpakdee2020empirical}, \emph{LIME} and \emph{BreakDown} were only examined on a single software defect prediction setting which leaves unanswered the more directly relevant question:
\textit{Are model-agnostic techniques reliable and stable under defect prediction models with different settings?} 
The answer to this question is critical.
{First, many studies conduct defect prediction under different settings.
The explanations generated of model-agnostic techniques are expected to be consistent across different settings to make them reliable and stable.}
{Second, we have seen many studies follow Jiarpakdee et al.~\cite{jiarpakdee2020empirical} to use model-agnostic techniques for other tasks, e.g., defective line prediction~\cite{wattanakriengkrai2020predicting}, online buggy commit prediction~\cite{pornprasit2021jitline}, and software quality assurance planning~\cite{rajapaksha2021sqaplanner}, understanding the reliability and stability of model-agnostic techniques will help confirm the findings from inline studies and benefit future research.}

In this study, we investigate the reliability and stability of model-agnostic techniques (i.e., \emph{LIME} and \emph{BreakDown}) on software defect prediction models under different settings. 
Specifically, we consider three different settings when building software defect prediction models, i.e., data sampling techniques, machine learning classifiers, and prediction scenarios.  
Data sampling techniques are used in software defect prediction studies~\cite{wang2016automatically,tan2015online,jiang2013personalized} to solve the data imbalance issue.
In this work, we experiment with five widely used sampling methods (details are in Section~\ref{sec:sampling}). 
Various machine learning classifiers, e.g., Logistic Regression (LR), Decision Tree (DT), and Random Forest (RF), etc., have been used to build defect prediction models~\cite{kamei2012large, lessmann2008benchmarking, gao2016software, zimmermann2009cross}. 
In this work, we experiment with {six} common machine learning classifiers (details are in Section~\ref{sec:classifiers}). 
Defect prediction includes two major scenarios, i.e., cross-version and cross-project defect prediction, in both scenarios, one can choose different versions of historical data to build the models.
In this work, we also examine the reliability and stability of \emph{LIME} and \emph{BreakDown} on these two scenarios when using different versions of data to build the defect prediction model. 

For our analysis, we reuse the same dataset from Jiarpakdee et al.~\cite{jiarpakdee2020empirical}, which contains 32 versions of defect data from nine large-scale open-source Java projects.
We run both \emph{LIME} and \emph{BreakDown} to generate explanations for the same instances under defect prediction models with different settings and then check the consistency of the generated explanations for the instances. 
Our experimental results show that explanations generated by both \emph{LIME} and \emph{BreakDown} are significantly inconsistent when different settings are applied, {which makes them unreliable to be used in practice.
In addition, our manual analysis shows that none of the generated explanations can reflect the root causes of the predicted defects, which further weakens the usefulness of model-agnostic based explanation generation.} 
Hence, {contrary to the claim of Jiarpakdee et al.~\cite{jiarpakdee2020empirical}, our study suggests 
that model-agnostic techniques 
are neither reliable nor stable to be used for explanation generation for defect prediction.} 
Overall, with this study, {we urge to revisit of other explainable software analytics studies that adopt model-agnostic techniques} and call for more research in explainable software defect prediction towards achieving consistent explanation generation.

This paper makes the following contributions:

\begin{itemize}
\item We perform the first study to analyze the reliability and stability of state-of-the-art model-agnostic based explanation generation techniques, i.e., \emph{LIME} and \emph{BreakDown} on software defect prediction. 

\item We examine the consistency of explanations generated by \emph{LIME} and \emph{BreakDown} under software defect prediction models with three typical settings, i.e., data sampling techniques, machine learning classifiers, and prediction scenarios.

\item We show neither \emph{LIME} nor \emph{BreakDown} can generate consistent explanations and the generated explanations cannot reflect the root causes of the predicted defects.
This makes model-agnostic techniques neither reliable nor stable to be used in practice.
Thus, we {urge a revisit of existing model-agnostic based studies in software engineering} and call for more research in building reliable and stable explanation generation for software analytics. 

\item We release the source code and the dataset of this work to help other researchers replicate and extend our study\footnote{https://doi.org/10.5281/zenodo.5425868}. 
\end{itemize}

We organized the rest of this paper as follows.
Section~\ref{sec:bm} presents
the background and motivation of this study. 
Section~\ref{sec:experiment} shows the experimental setup. 
Section~\ref{sec:result} presents the evaluation results. 
Section~\ref{sec:discussion} discusses open questions and the threats to the validity of this work. 
Section~\ref{sec:related} presents the related studies. 
Section~\ref{sec:conclusion} concludes this paper. 


%% file: sec/motivation.tex
\section{Background and Motivation}
\label{sec:bm}

This section introduces the background of software defect prediction models and the explanation generation techniques studied in this work and our motivation example. 

\subsection{File-level Defect Prediction Models}
\label{sec:dbmodel}
The objective of a file-level defect prediction model is to determine risky files for further software quality assurance activities. 
~\cite{hassan09icse,zimmermann2007predicting,rahman2013and,lee2011micro,moser2008comparative,kim2007predicting,nam2013transfer}. 
A typical release-based file-level defect prediction 
model mainly has three steps. 
The first step is to label the files in an early version as buggy or clean based on post-release defects for each file.  
Post-release defects are defined as defects that are revealed within a post-release window period (e.g., six months)~\cite{rahman2013and,yan2017file}.
One could collect these post-release defects from a Bug Tracking System 
(BTS) via linking bug reports to its bug-fixing changes. 
Files related to these bug-fixing changes are considered buggy.
Otherwise, the files are labeled as clean. 
The second step is to collect the corresponding defect features to represent these files.  
Instances with features and labels are used to train machine learning classifiers. 
Finally, trained models are used to predict files in a later version as buggy or clean. 

Following Jiarpakdee et al.~\cite{jiarpakdee2020empirical}, this paper also focuses on file-level defect prediction.
\input{figure/motivation}
\subsection{Model-agnostic based Explanation Generation Techniques}
\label{sec:expTech}
Model-agnostic techniques were originally introduced to explain the prediction of black-box AI/ML algorithms by identifying the contribution that each metric has on the prediction of an instance according to a trained model~\cite{ribeiro2016model}. 
\emph{LIME}~\cite{ribeiro2016should} and \emph{BreakDown}~\cite{gosiewska2019ibreakdown, staniak2018explanations} are two state-of-the-art model-agnostic explanation techniques.

\emph{LIME}~\cite{ribeiro2016should} mimics a black-box model it aims to explain.
To generate an explanation of an instance, \emph{LIME} follows four major steps.
It first creates synthetic instances around the instance to be explained.
Then, it generates predictions of all the synthetic instances generated in the step above.
After that, it creates a local regression model with the synthetic instances and their predictions made in the step above.
Finally, using the regression model, \emph{LIME} ranks the contribution of each metric to the predictions aligning with the black-box model. 
\emph{BreakDown}~\cite{gosiewska2019ibreakdown, staniak2018explanations} measures the additive contribution of each feature of an instance sequentially, summing up to the final black-box prediction result. 
In our study, we used the ag-break version of the \emph{BreakDown} technique, which works for non-additive models following Jiarpakdee et al.~\cite{jiarpakdee2020empirical}.

Jiarpakdee et al.~\cite{jiarpakdee2020empirical} are the first to leverage model-agnostic explanation techniques to generate 
instance explanations, which refer to an explanation of the prediction of defect prediction models.
The techniques define explanations as a list of ordered features. 
In this work, we empirically evaluate the reliability and stability of model-agnostic explanation techniques on software defect prediction models with different settings. 

\subsection{Motivating Example}
\label{sec:motivation}

In this section, we introduce an example to illustrate the problem of explanations generated by a model-agnostic technique, i.e., \emph{LIME}, which motivates us to further explore the reliability and stability of model-agnostic based explanation generation. 

Figure~\ref{fig:motiv_ex}
shows the explanations generated by \emph{LIME} for file \textit{``ActiveMQConnection.java''} with different software defect prediction models (i.e., LR in Figure~\ref{fig:motiv_ex_1} and DT in Figure~\ref{fig:motiv_ex_2}) from version 5.0.0 of project ActiveMQ. 
The figures list the ranking of features that contribute to the prediction, i.e., explanations of the prediction.
Figures on the left side are the probability and explanation of features that contribute to a prediction.
On the right side, the figure depicts the actual value of the feature.
For example, in Fig.~\ref{fig:motiv_ex_1}, ``COMM'' contributes 0.39 buggy-prone because the value is 11, which is over 3.
The orange color shows that a feature contributes in predicted as buggy and blue shows it contributes in predicted as clean. 
{Although under both software defect prediction models, the file is predicted as buggy, the generated explanations are significantly different.
Specifically, among the ten features selected by \emph{LIME}
on the LR-based defect prediction model, only two were also selected on the DT-based defect prediction model, i.e., ``DDEV'' and ``MaxCylomatic''.
However, ``DDEV'' and ``MaxCylomatic'' have different ranks.}
With this much difference, the generated explanation is unreliable and hard to be trusted. 

Motivated by this example, in this work we perform a comprehensive assessment and in-depth analysis of the state-of-the-art model-agnostic based explanation generation techniques, i.e., \emph{LIME} and \emph{BreakDown} on defect prediction models with different settings. {Note that the goal of this study is to evaluate the reliability and stability of a model-agnostic technique against itself under different software defect prediction models, not to evaluate one model-agnostic technique against another.}


%% file: figure/motivation.tex
\begin{figure*}[t!]
     \centering
     \begin{subfigure}[b]{0.49\textwidth}
         \centering
         \includegraphics[width=1\textwidth]{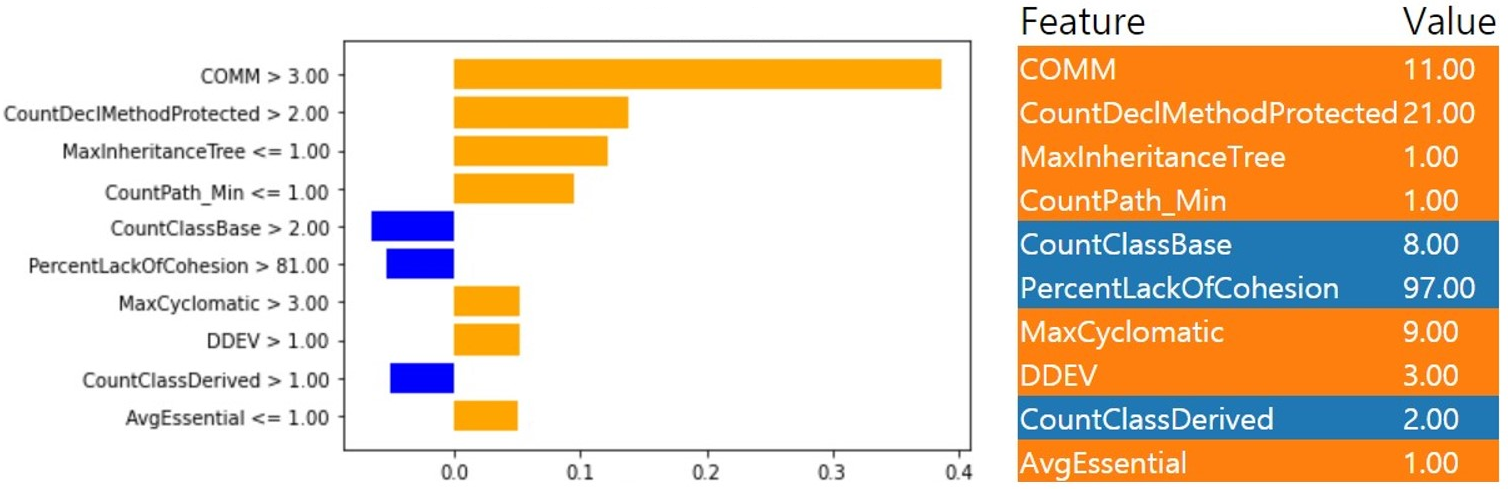}
         \caption{Explanation with LR-based defect prediction model} 
         \label{fig:motiv_ex_1}
     \end{subfigure}
     \begin{subfigure}[b]{0.49\textwidth}
         \centering
         \includegraphics[width=1\textwidth]{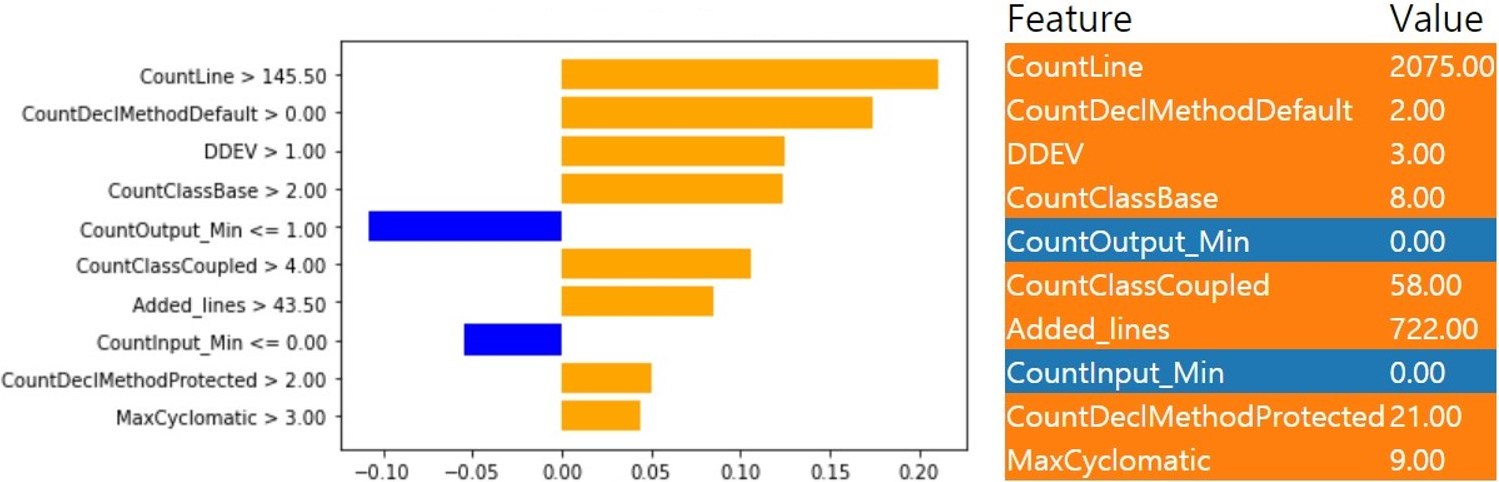}
         \caption{Explanation with DT-based defect prediction model}
         \label{fig:motiv_ex_2}
     \end{subfigure}
     \hfill
        \caption{
        Explanations generated by LIME for the predicted buggy file \textit{``ActiveMQConnection.java''} from project activemq-5.0.0 with different defect prediction models.}
        \label{fig:motiv_ex}
\end{figure*}

%% file: sec/setup.tex
\section{Empirical Study Setup}
\label{sec:experiment}

This section describes our experiment method for evaluating the reliability and stability of model-agnostic based explanation generation techniques, i.e., \emph{LIME} and \emph{BreakDown}, on defect prediction models in various settings.


\subsection{Research Questions}
\label{sec:rq}
To achieve the mentioned goal, we have designed experiments to answer the following research questions regarding the reliability and stability of each of the two studied model-agnostic based explanation generation techniques (i.e., \emph{LIME} and \emph{BreakDown}):
\vspace{4pt}

\noindent \textbf{RQ1:} \textit{Are the generated explanations from the same tool consistent under different data sampling techniques?} 

\vspace{4pt}

Software defect data are often imbalanced~\cite{tan2015online}, i.e., the buggy instances are much fewer than the clean ones.
Resampling, which changes the distribution between the majority class and the minority class, is an effective way to mitigate the effects of imbalanced data~\cite{wang2013using,wang2016automatically}. 
Thus, applying data sampling techniques is a common step in defect prediction~\cite{tan2015online}.
However, in Jiarpakdee et al.~\cite{jiarpakdee2020empirical}, no data sampling techniques were applied. 
In RQ1, we investigate whether a model-agnostic technique based explanation tool can generate consistent explanations for the predicted buggy instances under the software defect prediction model with applied different data sampling techniques. 

\vspace{4pt}
\noindent \textbf{RQ2:} \textit{Are the generated explanations from the same tool consistent under different machine learning classifiers?}  

\vspace{4pt}

To build accurate software defect prediction models, different machine learning classifiers have been used to build software defect prediction models, e.g., Logistic Regression (LR), Decision Tree (DT), Random Forest (RF), etc.
Researchers have examined that these machine learning classifiers achieve the best performance on different datasets or under different prediction scenarios~\cite{kamei2012large, lessmann2008benchmarking, gao2016software, zimmermann2009cross}.
In RQ2, we examine whether model-agnostic based explanation tools can generate consistent explanations for the predicted buggy instances under the software defect prediction model with applied different machine learning classifiers. 

\vspace{4pt}
\noindent \textbf{RQ3:} \textit{Are the generated explanations from the same tool consistent under cross-version defect prediction scenarios?}

\vspace{4pt}

Cross-version defect prediction is one type of within-project defect prediction~\cite{turhan2009relative}, which is often used for projects that have sufficient history data, e.g., a project can have multiple releases.
One can choose different history versions as the training data to build the models with different training data selection approaches~\cite{wang2021continuous}.
In Jiarpakdee et al.~\cite{jiarpakdee2020empirical}, the defect prediction model was trained and tested on data from the same version (i.e., within-version defect prediction).
In RQ3, we explore whether a model-agnostic based explanation tool can generate consistent explanations for the predicted buggy instances under the software defect prediction models trained on different history releases from the same project.

\vspace{4pt}
\noindent \textbf{RQ4:} \textit{Are the generated explanations from the same tool  consistent under cross-project prediction scenarios?}

\vspace{4pt}

In cross-project defect prediction (CPDP), the training and test datasets are from different projects.
CPDP is designed for projects that do not have historical data~\cite{nam2013transfer}.
Jiarpakdee et al.~\cite{jiarpakdee2020empirical} did not examine the performance of model-agnostic based explanation in a cross-project scenario. 
In RQ4, we explore whether a model-agnostic based explanation tool can generate consistent explanations for the predicted buggy instances under the software defect prediction model trained on data from different projects. 



\subsection{Experiment Data}
\label{sec:dataset}
\input{table/projs}
\input{table/features}

In this paper, to avoid potential bias introduced by experiment data, we reuse the same defect data from Jiarpakdee et al.~\cite{jiarpakdee2020empirical}, which comprises 32 releases that span 9 open-source software systems.
Table~\ref{tab:expprojects} shows the statistical information of the dataset.

For building defect prediction models, we also reuse the same software metrics used in Jiarpakdee et al.~\cite{jiarpakdee2020empirical}. In total, 65 software metrics along 3 dimensions are used, i.e., 54 code metrics (describe the relationship between properties extracted from source code and software quality), 5 process metrics (describe the relationship between development
process and software quality), and 6 human metrics (describe the relationship between the ownership of instances and software quality).
Table~\ref{tab:metrics} shows the metrics used to build defect prediction models in this work.
Note that, Jiarpakdee et  al.~\cite{jiarpakdee2020empirical} has applied AutoSpearman~\cite{jiarpakdee2018autospearman} to remove irrelevant metrics and correlated metrics before experiments.
As a result, only 22-27 of the 65 metrics were used in the experiments.
We follow the same process in this study to avoid any potential bias introduced by data pre-processing.

\subsection{Studied Data Sampling Techniques}
\label{sec:sampling}
In this study, we examine the consistency of an explanation generation tool under five widely used data sampling methods, which are shown as follows. 
\begin{itemize}
    \item \textbf{Cluster Centroids~\cite{altinccay2004clustering}}: performs an under-sampling by using centroids as the new majority samples made by k-means clusters. 
     \item \textbf{Repeated Edited Nearest Neighbours (RENN)~\cite{tomek1976experiment}}: applies the nearest-neighbour algorithm to edit the samples by removing instances that are not similar to their neighbours.
     \item \textbf{Random under-sampling (RUS)~\cite{bach2019proposal, LARADJI2015388, wahono2015systematic}}: randomly picks samples from the majority class to match the minority class.
    \item \textbf{Random over-sampling (ROS)~\cite{ling1998data, LARADJI2015388}}: over-samples the minority class by picking random samples with replacement.
    \item \textbf{SMOTE~\cite{chawla2002smote, LARADJI2015388}}: is the synthetic minority over-sampling technique (\emph{SMOTE}).
    This method creates synthetic examples of the minority class rather than over-sampling with replacements.

\end{itemize}
Researchers have widely used all the above data sampling techniques in software prediction tasks~ \cite{rodriguez2014preliminary,pak2018empirical,feng2021investigation,chen2018tackling, wang2018top,wang2013using,li2012sample}. 
In this work, we use the implementations of these data sampling techniques from the widely used imbalanced-learn Python library~\cite{imblearn}.

\subsection{Studied Defect Prediction Classifiers}
\label{sec:classifiers}

Jiarpakdee et al.~\cite{jiarpakdee2020empirical} showed that the model-agnostic techniques can be applied to many machine learning classifiers for explanation generation tasks.
In this study, we use the same six machine learning classifiers mentioned in Jiarpakdee et al.~\cite{jiarpakdee2020empirical} to build the defect prediction models. The details of these classifiers are as follows: 
\begin{itemize}
    \item \textbf{Logistic Regression}: is the baseline model used in Jiarpakdee et al.~\cite{jiarpakdee2020empirical}. 
    It is a statistical model that uses the logistic function for classifying binary dependent variables. 
    Logistic Regression is still widely used in defect prediction due to its advanced performance despite its simplicity \cite{ghotra2015revisiting}.
    \item \textbf{Decision Tree (DT):} is a model that uses trees to observe which features affect a target class.
    \item \textbf{Random Forest (RF):} utilizes ensemble learning, which is a technique that combines many classifiers to provide solutions to complex problems. A random forest algorithm consists of many decision trees and makes decisions via majority voting of multiple decision trees. 
    \item \textbf{Averaged Neural Network (AVNNet):} is a neural network model that uses models with different random numbers as seeds. It averages all the resulting models to make a prediction decision. 
    \item \textbf{Gradient Boosting Machine (GBM):} uses an additive model of a forward stage-wise fashion. It uses predictors, such as decision trees, to form an ensemble.  
    \item \textbf{Extreme Gradient Boosting Tree (xGBTree):} follows the gradient boosting principle. However, xGBTree uses a more regularized model to control over-fitting efficiently. 
\end{itemize}
In this work, we used the implementation of the above six machine learning classifiers developed in the scikit-learn library~\cite{scikit_learn} and xgboost\footnote{\url{https://xgboost.readthedocs.io/en/latest/python/index.html}}.
Note that we have also tuned each of the six classifiers with its parameters and use the ones that can achieve the best AUC values to build prediction models in our experiments, as suggested in~\cite{jiarpakdee2020empirical}. 

\subsection{Studied Defect Prediction Scenarios}
Software defect prediction models can be categorized into within-project and cross-project models based on the source of the training and test datasets.
In within-project defect prediction, one can choose different history versions as the training data to build the models, which we call cross-version defect prediction. 
In this study, we investigate the consistency of explanations generated by the same tool under the two following defect prediction scenarios.

\subsubsection{Cross-Version Defect Prediction}
The Cross-version defect prediction scenario is one of the actively studied scenarios in within-project defect prediction~\cite{xu2018cross, yang2018ridge, amasaki2020cross}.
In this paper, to perform a cross-version defect prediction scenario, for each project, we use its latest version as the test version and randomly select two earlier versions as the training data to build defect prediction models respectively. 


\subsubsection{Cross-project Defect Prediction}

The cross-project defect prediction scenario is also another actively studied scenario in the defect prediction field~\cite{zimmermann2009cross, he2012investigation, guo2016cold}. 
To perform cross-version defect prediction, we randomly select one version from each project as the test dataset and then we randomly select two different versions from two different projects as the training data to build software defect prediction models, respectively. 

For both cross-version and cross-project defect prediction, given a test dataset, we use two different defect models to predict bugs on it, and then we run a model-agnostic based explanation generation technique to generate explanations on files that are predicted as buggy under both models to check whether the generated explanations of the same tool are consistent.  
We iterate the random selection 10 times to avoid potential bias and report the average of the results. 
\subsection{Evaluation Measures} 
\label{sec:metric}

In this work, given a model-agnostic based explanation generation technique (i.e., \emph{LIME} and \emph{BreakDown}), we use the following two metrics to evaluate the consistency of two explanations generated by it under two different defect prediction models. They are 
{\hitrate} and {\rankdiff}.

$Hit\_rate$ is the percentage of features that match between the two explanations (i.e., a set of ranked features).
For instance, Jiarpakdee et al.~\cite{jiarpakdee2020empirical} leveraged the top-10 features 
ranked by model-agnostic techniques as the explanation to interpret the prediction results.
If $N$ ($N>=0$ and $N<=10$) out of the ten features are found in two explanations generated under two different software defect prediction models, the value of {\hitrate} between these two explanations is $\frac{N}{10}$.
$Hit\_rate$ indicates how similar the two explanations are without considering the ranking orders of features in the explanations. 
The range of the {\hitrate} value is from 0.0 to 1.0.
The higher the {\hitrate}, the better the consistency of an explanation generation technique is. 
In our experiments, we use the top 10 features for \emph{LIME} and \emph{BreakDown} to calculate {\hitrate} as used in Jiarpakdee et al.~\cite{jiarpakdee2020empirical}. 

Note that, since {\hitrate} does not consider the order of features in the explanations, 
we also introduce {\Rankdiff}, which compares two explanations by using the orders of features in the explanations. 
Specifically, {\Rankdiff} measures the average difference of feature rankings between two explanations. 
For instance, if a feature is ranked $M$th and $H$th in two different explanations, the ranking difference of it is $ abs\left (M - H  \right )$. 
{\Rankdiff} is reported as the average ranking difference of all features in two explanations.
If a feature is not in the ranking, the difference is set to top-N. 
Higher {\rankdiff} means more different the explanations are by rankings.
The range of the {\rankdiff} is from zero (all features match all ranking orders) to the number of total features considered in the explanation, i.e., 10 (all features don't appear in the top 10). 
The smaller the {\rankdiff}, the better the consistency of an explanation generation technique is.



%% file: table/projs.tex
\begin{table}[t!]
\centering
\caption{Subjects studied in this work} 
\label{tab:expprojects}
\scalebox{0.95}{
\setlength\tabcolsep{3pt} 
\begin{tabular}{l|c|c|c|l}
\hline
\textbf{Project}  &\textbf{\#Files} & \textbf{\#KLOC} & \textbf{Bug rate} & \textbf{Studied Releases}  \\ \hline
ActiveMQ &  1.8K-3.4K  & 142-299 &  6\%-15\% & 5.0,5.1,5.2,5.3,5.8  \\ \hline
Camel &   1.5K-8.8K &  75-383 &2\%-18\% &1.4,2.9,2.10,2.11  \\ \hline
Derby & 1.9K-2.7K &  412-533 &14\%-33\% &10.2,10.3,10.5  \\ \hline
Groovy & 0.7K-0.9K&74-90 & 3\%-8\% & 1.5.7,1.6.0.b1,1.6.0.b2  \\ \hline
HBase & 10K-18K &  246-534  &  20\%-26\% & 0.94,0.95.0,0.95.2 \\ \hline
Hive &   14K-27K  & 287-563  &  8\%-19\% & 0.9,0.10,0.12 \\ \hline
JRuby &   0.7K-16K  & 105-238 & 5\%-18\% & 1.1,1.4,1.5,1.7 \\ \hline
Lucene &  0.8K-28K  &   101-342 &  3\%-24\% & 2.3,2.9,3.0,3.1 \\ \hline
Wicket &  16K-28K  &109-165  & 4\%-7\% & 1.3.b1,1.3.b2,1.5.3 \\ \hline
\end{tabular}
}
\end{table}

%% file: table/features.tex
\begin{table*}[t]
	\centering
	\caption{Details of the metrics used to build software bug prediction models in this work.}
	\label{tab:metrics}
	\scalebox{0.9}{
\setlength\tabcolsep{4pt} 
\begin{tabular}{l|l|l}
\hline
\textbf{Type} & \textbf{Metrics} & \textbf{Count} \\ \hline
File   &  \begin{tabular}[c]{@{}l@{}}AvgCyclomatic, AvgCyclomaticModified, AvgCyclomaticStrict, AvgEssential, AvgLine, AvgLineBlank, AvgLineCode, AvgLineComment, \\ CountDeclClass, CountDeclClassMethod, CountDeclClassVariable, CountDeclFunction, CountDeclInstanceMethod,\\ CountDeclInstanceVariable, CountDeclMethod, CountDeclMethodDefault, CountDeclMethodPrivate, CountDeclMethodProtected,\\ CountDeclMethodPublic, CountLine, CountLineBlank, CountLineCode, CountLineCodeDecl, CountLineCodeExe, CountLineComment,\\ CountSemicolon, CountStmt, CountStmtDecl, CountStmtExe, MaxCyclomatic, MaxCyclomaticModified, MaxCyclomaticStrict, \\ RatioCommentToCode, SumCyclomatic, SumCyclomaticModified, SumCyclomaticStrict, SumEssential\end{tabular} & 37 \\ \hline
Class  & CountClassBase, CountClassCoupled, CountClassDerived, MaxInheritanceTree, PercentLackOfCohesion& 5 \\ \hline
Method & CountInput\_\{Min, Mean, Max\}, CountOutput\_\{Min, Mean, Max\}, CountPath\_\{Min, Mean, Max\}, MaxNesting\_\{Min, Mean, Max\}& 12 \\ \hline
Process & COMM, ADDED\_LINES, DEL\_LINES, ADEV, DDEV & 5 \\ \hline
Ownership & MINOR\_COMMIT, MINOR\_LINE, MAJOR\_COMMIT, MAJOR\_LINE, OWN\_COMMIT, OWN\_LINE & 6 \\ \hline
\end{tabular}
}
\end{table*}

%% file: sec/result.tex
\section{Results and Analysis}
\label{sec:result}
This section presents the experimental results and the answers to the research questions regarding the reliability and stability of model-agnostic techniques proposed in Section~\ref{sec:rq}.
\input{table/average}

\input{figure/rq2/rq2_combine}
\subsection{RQ1: Explanation Consistency Under Different Data Sampling Techniques}
\label{sec:answer_rq2}
\noindent \textbf{Approach:} To investigate the consistency of the generated explanations of a model-agnostic technique under different data sampling approaches, we combine the defect prediction setting (i.e., each of the classifiers listed in Section~\ref{sec:classifiers} in the within-version defect prediction scenario
used in Jiarpakdee et al.~\cite{jiarpakdee2020empirical} with different data sampling techniques to build defect prediction models on each experimental project. 
We follow Jiarpakdee et al. to use the out-of-sample bootstrap validation 
technique to create the training and test data on each version of each project listed in Table~\ref{tab:expprojects}.  
On the same test dataset, we run both \emph{LIME} and \emph{BreakDown} under Jiarpakdee et al.'s defect prediction model and its variant with data sampling techniques to generate explanations for test instances that are predicted as buggy in both models. 
We use the {\hitrate} and {\rankdiff} to evaluate the consistency of explanations generated by \emph{LIME} and  \emph{BreakDown}. 
{In total, we have 60 runs on each project, i.e., 6 classifiers * 5 data sampling * 2 options (with or without sampling), for both \emph{LIME} and \emph{BreakDown}.} 
We report the average values of {\hitrate} and {\rankdiff} of explanations generated by the same model-agnostic technique under defect prediction models {with or without} different data sampling techniques applied. In this RQ, we examine two typical model-agnostic techniques, i.e.,
\emph{LIME} and \emph{BreakDown}.\vspace{4pt}


\noindent \textbf{Result:} Table~\ref{tab:average} shows the average {\hitrate} and {\rankdiff} {of explanations generated from the same model-agnostic technique} before and after applying each data sampling technique. Figure~\ref{fig:rq2_comb} shows the detailed distribution of {\hitrate} and {\rankdiff} on each project. 
Overall, both \emph{LIME} and \emph{BreakDown} generate inconsistent explanations on defect prediction models before and after applying different data sampling techniques. 
Specifically, the average {\hitrate} values of \emph{LIME} and \emph{BreakDown} range from 0.574 (using Random under-sampling) to 0.641 (using Random over-sampling) and 0.655 (using Cluster Centroids) to 0.769 (using Random over-sampling), respectively,
which implies almost 40\% and 29\% of the features in the generated explanations of \emph{LIME} and \emph{BreakDown} are different before and after data sampling techniques are applied. 

Regarding {\rankdiff}, on average, 5 out of the 10 features in the explanations from \emph{LIME} and 4 out of 10 features from \emph{BreakDown} have different ranks,  
which implies on average 50\% and 40\% features in the explanations generated by \emph{LIME} and \emph{BreakDown} have a different order under defect prediction models before and after data sampling applied. 

In addition, we have also checked that for both \emph{LIME} and \emph{BreakDown}, {100\%} of test instances have different feature orders before and after applying data sampling techniques. 
From these observations, we can see that explanations generated by \emph{LIME} and \emph{BreakDown} are inconsistent when data sampling is applied, which makes them unreliable and unstable.  


\mybox{Both \emph{LIME} and \emph{BreakDown} generate inconsistent explanations when data sampling is applied. On average, almost 40\% of the features in the explanations generated by \emph{LIME} and 29\% from \emph{BreakDown} are different when data sampling techniques are applied. 
In addition, around 50\% and 40\% of features in the explanations generated by \emph{LIME} and \emph{BreakDown} have different orders under any data sampling technique.}


\subsection{RQ2: Explanation Consistency Under Different Classifiers}
\label{sec:answer_rq3}
\noindent \textbf{Approach:} To investigate the consistency of the generated explanations of the same model-agnostic technique under defect prediction models trained on different machine learning classifiers, 
we use the six widely-used machine learning classifiers as our experiment subjects (details are in Section~\ref{sec:classifiers}). 
Note that, to avoid potential bias, we do not apply any data sampling technique in RQ2. 
For each classifier, we follow the process described in Jiarpakdee et al.~\cite{jiarpakdee2020empirical} to create the training and test data. We use the LR-based software defect prediction model as the baseline as suggested in~\cite{jiarpakdee2020empirical} 
for the comparison. On the same test dataset, we run a model-agnostic technique on both the baseline (LR-based defect prediction model) and each of the other five examined classifiers, i.e., DT, RF, AVNNet, GBM, and xGBTree, to generate explanations for test instances. When different machine learning models are applied, prediction results of the same instance vary as buggy or clean.
So, we only consider instances that are predicted as buggy in both compared machine learning predictors. 
To measure the consistency, we use {\hitrate} and {\rankdiff} to evaluate \emph{LIME} and  \emph{BreakDown} on different classifiers. 
We report the average values of {\hitrate} and {\rankdiff} across all the experiment projects when comparing two classifiers. In this RQ, we also examine two model-agnostic techniques, i.e., \emph{LIME} and \emph{BreakDown}. 

\input{table/classifierAvg}

\vspace{4pt}

\noindent \textbf{Result:} 
Table~\ref{tab:ml_avg} shows the average {\hitrate} and {\rankdiff} of the two explanation generation tools on different machine learning classifiers. Overall, both \emph{LIME} and \emph{BreakDown} generate inconsistent explanations between different machine learning classifiers. 
For \emph{LIME}, the average {\hitrate} on these projects ranges from 0.515 (i.e., DT) to 0.613 (i.e., AVNNet), 
which means around 44\% of the features in \emph{LIME}'s explanations are different when a different machine learning classifier is applied for defect prediction compared to LR based defect prediction model. 
\emph{BreakDown} has a slightly higher {\hitrate}, around 36\% of the features in \emph{BreakDown}'s explanations are different when different machine learning classifiers are applied. 
In addition, all the {\rankdiff} values of \emph{LIME} and \emph{BreakDown} are higher than 4 and our analysis further reveals that on average there are more than 5 and 4 features in the explanations generated by \emph{LIME} and  \emph{BreakDown} that have different ranks under software defect prediction models with different classifiers, which indicates 50\% and 40\% features in the generated explanations have different orders. 
Note that, because of the space limitation, we only show the results of experiments whose base model is LR, we have also used each of the studied machine learning classifiers as the base model, and we observe similar findings, which indicates \emph{LIME} and \emph{BreakDown} consistently generate unreliable explanations when different classifiers are applied.

\mybox{
Both \emph{LIME} and \emph{BreakDown} generate inconsistent explanations under different classifiers. 
Specifically, on average, 44\%
of the features in \emph{LIME}’s explanations and 36\% of the features in \emph{BreakDown}’s explanations are different when different machine learning classifiers are applied. 
In addition, more than 50\% and 40\% of the features in the explanations generated by \emph{LIME} and \emph{BreakDown} have different orders when different machine learning classifiers applied.} 

\input{figure/rq4n5/rq4_combine}
\input{figure/rq4n5/rq5_combine}
\subsection{RQ3: Explanation Consistency Under the Cross-Version Scenario}
\label{sec:answer_rq4}

\noindent \textbf{Approach:} To investigate the consistency of the generated explanations of a model-agnostic technique under cross-version defect prediction scenario, for each experiment project listed in Table~\ref{tab:expprojects}, we use its latest version as the test data, and we then randomly select two different versions from the same project as the training data to train two different software defect prediction models.
We run the model-agnostic technique under both models to generate explanations for test instances that are predicted as buggy in both models. 
We use the {\hitrate} and {\rankdiff} to evaluate the consistency of explanations generated by the model-agnostic technique. 
Note that, in this study, we use six different classifiers (details are in Section~\ref{sec:classifiers}) and examine two model-agnostic techniques, i.e., \emph{LIME} and \emph{BreakDown}. 

\vspace{4pt}

\noindent \textbf{Result:} 
Table~\ref{tab:rq34} shows the average {\hitrate} and {\rankdiff} of \emph{LIME} and \emph{BreakDown} under the cross-version prediction scenario and Figure~\ref{fig:rq4_comb} shows the detailed distributions of {\hitrate} and {\rankdiff}.  
As we can see from the results, the {\hitrate} values of both \emph{LIME} and \emph{BreakDown} are higher than 0.4 on each project. 
On average, {\hitrate} is 0.518 across all the projects for \emph{LIME}, which means around 50\% of the generated explanations of \emph{LIME} are different under cross-version defect prediction. 
For \emph{BreakDown}, we can see that its average {\hitrate} is 0.591, indicating 41\% of the generated explanations are different under cross-version defect prediction. 

In addition, we can see that the {\rankdiff} of both \emph{LIME} and \emph{BreakDown} on most projects are higher than 5, which indicates around 50\% of features in the generated explanations of both \emph{LIME} and \emph{BreakDown} have different orders under defect prediction models built on different versions. 


\mybox{Both \emph{LIME} and \emph{BreakDown} generate inconsistent explanations under cross-version defect prediction scenarios. 
Overall, 50\% of features in the generated explanations of \emph{LIME} and 41\% of \emph{BreakDown} are different. In addition, around 50\% of features in the generated explanations of \emph{LIME} and \emph{BreakDown} have different orders under the cross-version defect prediction scenario.} 
\input{table/RQ34}

\subsection{RQ4: Explanation Consistency Under the Cross-Project Scenario}
\label{sec:answer_rq5}

\noindent \textbf{Approach:} To investigate the consistency of the explanations generated by a model-agnostic technique under cross-project defect prediction scenario, we first randomly select one version from each experiment project as the test data, 
we then randomly select two different versions from two different projects respectively as the training data to build two defect prediction models.
We run the model-agnostic technique to generated explanations for test instances that are predicted as buggy in both models. 
We then use the {\hitrate} and {\rankdiff} to evaluate the consistency of explanations generated by the model-agnostic technique.  
In addition, we use each of the six studied classifiers to run the experiments. 
We repeat the above process 10 times for avoiding potential data selection bias. 
Thus, each project has 6*10 experiment runs for examining the consistency of the generated explanations. 
We examine the studied two model-agnostic techniques, i.e., \emph{LIME} and \emph{BreakDown}. 
Similar to other RQs, we use the average {\hitrate} and {\rankdiff} of each run to measure the consistency. 


\input{figure/diff_n/diff_n}

\vspace{4pt}

\noindent \textbf{Result:} 
Table~\ref{tab:rq34} shows the average {\hitrate} and {\rankdiff} of the two explanation generation tools under the cross-project prediction scenario. 
Figure~\ref{fig:rq5_comb} presents the detailed distribution of {\hitrate} and {\rankdiff} for \emph{LIME} and \emph{BreakDown} on each project. As we can see from the figures, the average {\hitrate} of \emph{LIME} is around 0.48 on each project, which meaning around 52\% of features in the generated explanations of \emph{LIME} are different under the cross-project defect prediction scenario. 
For \emph{BreakDown}, the average {\hitrate} is 0.51, indicating 49\% features in the generated explanations are different under the cross-project defect prediction scenario. 

In addition, we can see that the {\rankdiff} values both of \emph{LIME} and \emph{BreakDown} are around 6 on each project, which indicates 60\% of features in the generated explanations of them have different orders under the cross-project defect prediction scenario.


\mybox{Both \emph{LIME} and \emph{BreakDown} generate inconsistent explanations under the cross-project defect prediction scenario. 
Specifically, around 52\% of the features in the generated explanations of \emph{LIME} and 49\% for \emph{BreakDown} are different under the cross-project defect prediction scenario. In addition, 60\% of features in the generated explanations have different orders.}  

\input{table/discussion2}

%% file: table/average.tex

\begin{table}[t!]
\centering
\caption{Average {\hitrate} and {\rd} of the explanations generated by \emph{LIME} and \emph{BreakDown} under defect prediction models before and after applying different data sampling techniques.}
\label{tab:average}
\setlength\tabcolsep{3.0pt} 
\scalebox{0.95}{
\begin{tabular}{|c|c|c|c|c|}
\hline
\multicolumn{1}{|c|}{\multirow{2}{*}{\textbf{\begin{tabular}[c]{@{}c@{}}Data sampling \\ techniques\end{tabular}}}} & \multicolumn{2}{c|}{\textbf{LIME}} & \multicolumn{2}{c|}{\textbf{BreakDown}} \\ \cline{2-5} 
\multicolumn{1}{|c|}{} & \hitrate & \rd & \hitrate & \rd \\ \hline
Cluster Centroids & 0.577 & 5.564 & 0.655 & 4.725 \\ \hline
Repeated Edited NN & 0.608 & 5.328 & 0.692 & 4.292 \\ \hline
Random under-sampling & 0.574 & 5.655 & 0.686 & 4.382\\ \hline
Random over-sampling & 0.641 & 4.892 & 0.769 & 3.505\\ \hline
SMOTE & 0.632 & 4.948 & 0.762 & 3.577\\ \hline \hline
\textbf{Average} &0.606&5.277&0.713&4.096\\ \hline
\end{tabular}
}
\end{table}

%% file: figure/rq2/rq2_combine.tex
\begin{figure*}[th!]

\centering 
\begin{subfigure}{0.20\linewidth}
\centering
\includegraphics[width=\textwidth]{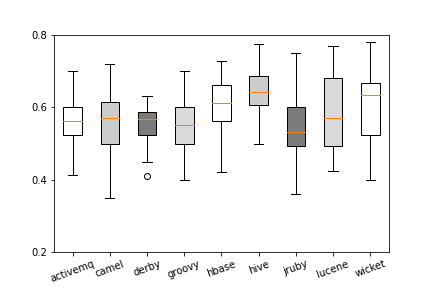}
\vspace{-0.05in}
\caption{Cluster Centroids}
\end{subfigure}\hfill
\begin{subfigure}{0.20\linewidth}
\centering
\includegraphics[width=\textwidth]{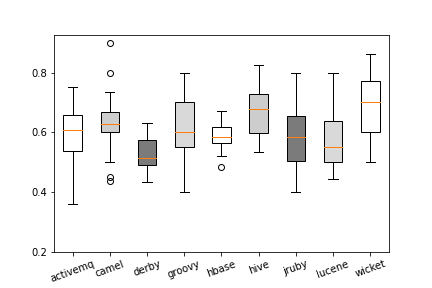}
\vspace{-0.05in}
\caption{RENN}
\end{subfigure}\hfill
\begin{subfigure}{0.20\linewidth}
\centering
\includegraphics[width=\textwidth]{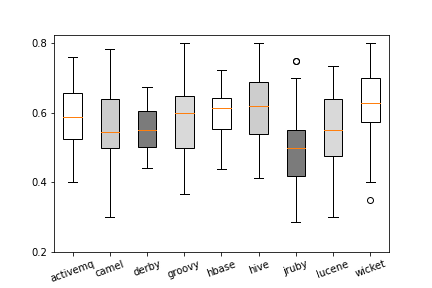}
\vspace{-0.05in}
\caption{RUS}
\end{subfigure}\hfill
\centering
\begin{subfigure}{0.20\linewidth}
\centering
\includegraphics[width=\textwidth]{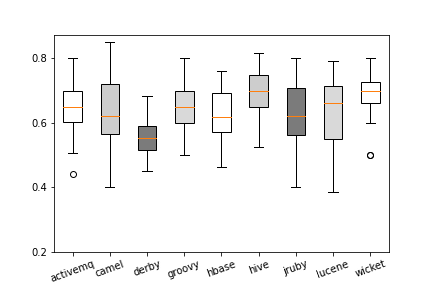}
\vspace{-0.05in}
\caption{ROS}
\end{subfigure}\hfill
\centering
\begin{subfigure}{0.20\linewidth}
\centering
\includegraphics[width=\textwidth]{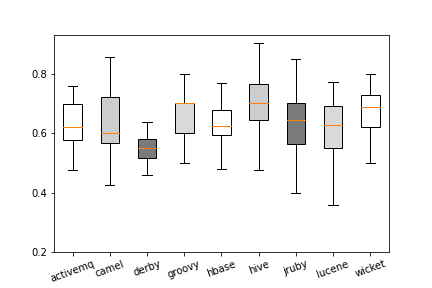}
\vspace{-0.05in}
\caption{SMOTE}
\end{subfigure}\hfill

\centering 
\begin{subfigure}{0.20\linewidth}
\centering
\includegraphics[width=\textwidth]{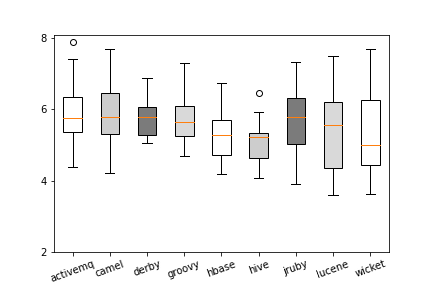}
\vspace{-0.05in}
\caption{Cluster Centroids}
\end{subfigure}\hfill
\begin{subfigure}{0.20\linewidth}
\centering
\includegraphics[width=\textwidth]{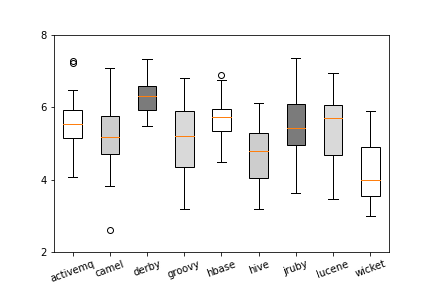}
\vspace{-0.05in}
\caption{RENN}
\end{subfigure}\hfill
\begin{subfigure}{0.20\linewidth}
\centering
\includegraphics[width=\textwidth]{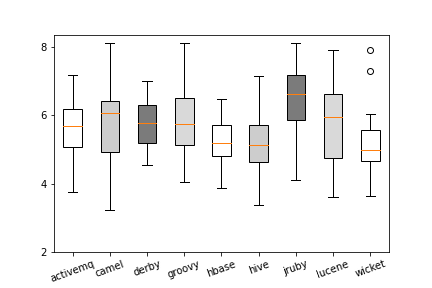}
\vspace{-0.05in}
\caption{RUS}
\end{subfigure}\hfill
\centering
\begin{subfigure}{0.20\linewidth}
\centering
\includegraphics[width=\textwidth]{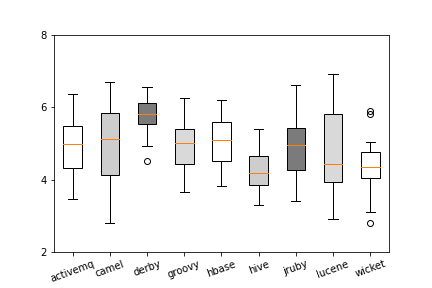}
\vspace{-0.05in}
\caption{ROS}
\end{subfigure}\hfill
\centering
\begin{subfigure}{0.20\linewidth}
\centering
\includegraphics[width=\textwidth]{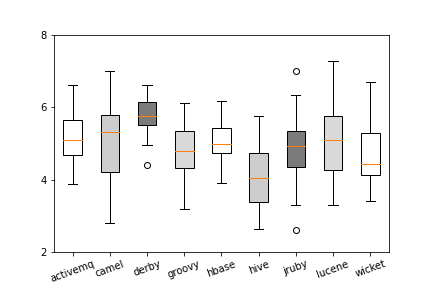}
\vspace{-0.05in}
\caption{SMOTE}
\end{subfigure}\hfill

\centering 
\begin{subfigure}{0.20\linewidth}
\centering
\includegraphics[width=\textwidth]{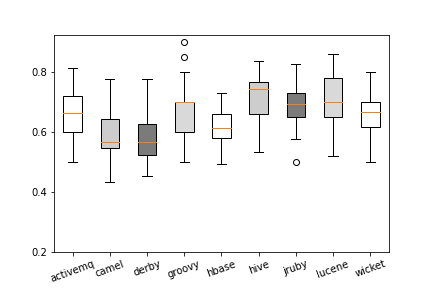}
\vspace{-0.05in}
\caption{Cluster Centroids}
\end{subfigure}\hfill
\begin{subfigure}{0.20\linewidth}
\centering
\includegraphics[width=\textwidth]{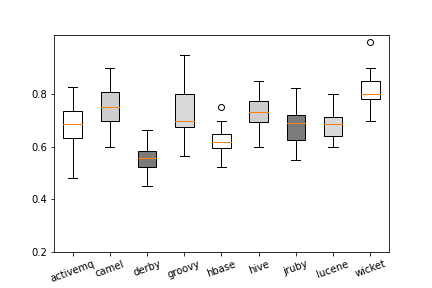}
\vspace{-0.05in}
\caption{RENN}
\end{subfigure}\hfill
\begin{subfigure}{0.20\linewidth}
\centering
\includegraphics[width=\textwidth]{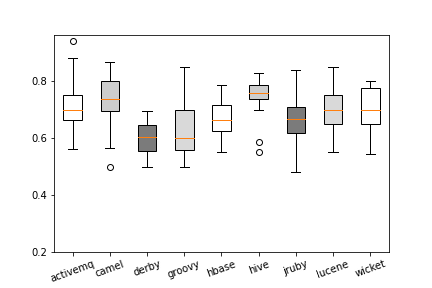}
\vspace{-0.05in}
\caption{RUS}
\end{subfigure}\hfill
\centering
\begin{subfigure}{0.20\linewidth}
\centering
\includegraphics[width=\textwidth]{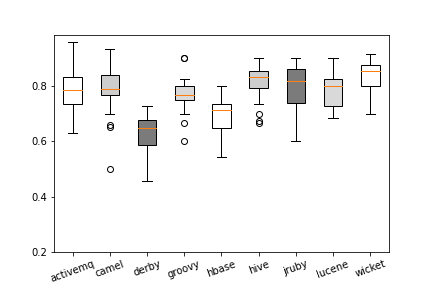}
\vspace{-0.05in}
\caption{ROS}
\end{subfigure}\hfill
\centering
\begin{subfigure}{0.20\linewidth}
\centering
\includegraphics[width=\textwidth]{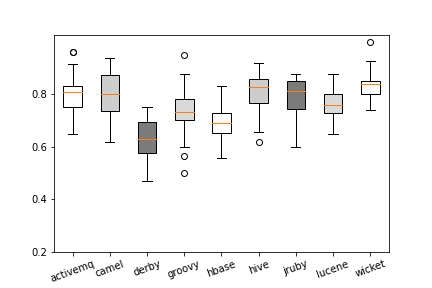}
\vspace{-0.05in}
\caption{SMOTE}
\end{subfigure}\hfill

\centering 
\begin{subfigure}{0.20\linewidth}
\centering
\includegraphics[width=\textwidth]{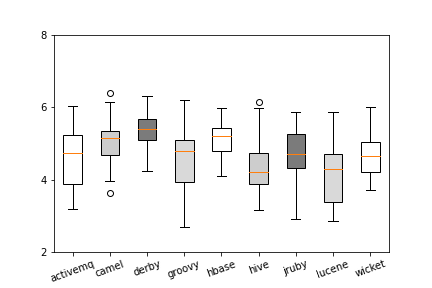}
\vspace{-0.05in}
\caption{Cluster Centroids}
\end{subfigure}\hfill
\begin{subfigure}{0.20\linewidth}
\centering
\includegraphics[width=\textwidth]{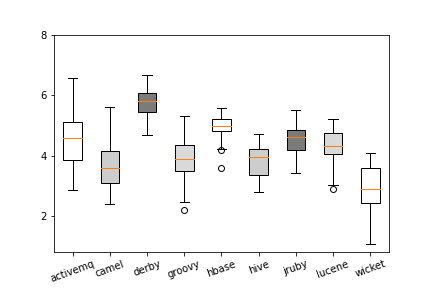}
\vspace{-0.05in}
\caption{RENN}
\end{subfigure}\hfill
\begin{subfigure}{0.20\linewidth}
\centering
\includegraphics[width=\textwidth]{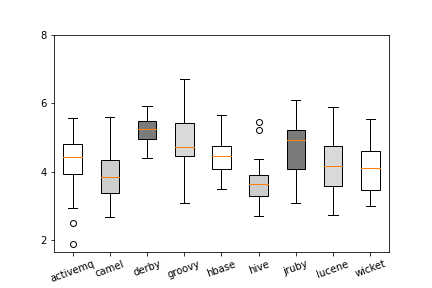}
\vspace{-0.05in}
\caption{RUS}
\end{subfigure}\hfill
\centering
\begin{subfigure}{0.20\linewidth}
\centering
\includegraphics[width=\textwidth]{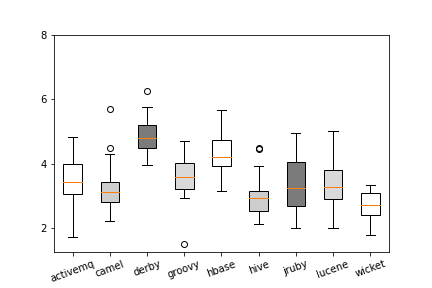}
\vspace{-0.05in}
\caption{ROS}
\end{subfigure}\hfill
\centering
\begin{subfigure}{0.20\linewidth}
\centering
\includegraphics[width=\textwidth]{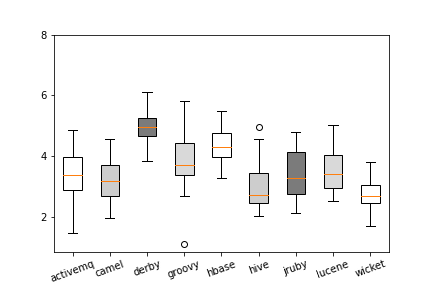}
\vspace{-0.05in}
\caption{SMOTE}
\end{subfigure}\hfill
\caption{{The distributions of {\hitrate} and {\rd} on each project before and after applying data sampling techniques. 
(a) $\sim$ (e) is the result of {\hitrate} of \emph{LIME}.
(f) $\sim$ (j) is the result of {\rd} of \emph{LIME}.
(k) $\sim$ (o) is the result of {\hitrate} of \emph{BreakDown}.
(p) $\sim$ (t) is the result of {\rd} of \emph{BreakDown}.}}
\label{fig:rq2_comb}
\end{figure*}

%% file: table/classifierAvg.tex
\begin{table}[t!]
\centering
\caption{Average {\hitrate} and {\rd} of the explanations generated by \emph{LIME} and \emph{BreakDown} under  defect prediction models with different classifiers compared to the LR-based defect prediction model.
}
\label{tab:ml_avg}
\setlength\tabcolsep{3.0pt} 
\scalebox{1}{
\begin{tabular}{|l|c|c|c|c|}
\hline
\multirow{2}{*}{\textbf{Classifier}} & \multicolumn{2}{c|}{\textbf{LIME}} & \multicolumn{2}{c|}{\textbf{BreakDown}} \\ \cline{2-5} 
  & {\hitrate} & {\rd} & {\hitrate} & {\rd} \\ \hline
AVNNet & 0.613 & 5.325 & 0.681 & 4.339 \\ \hline
DT & 0.515 & 6.185 & 0.609 & 5.241 \\ \hline
GBM  & 0.559 & 5.714 & 0.638 & 4.826 \\ \hline
RF & 0.557 & 5.712 & 0.649 & 4.739 \\ \hline
XGB  & 0.570 & 5.564 & 0.641 & 4.778 \\ \hline \hline
\textbf{Average} &0.563&5.700&0.644& 4.785\\ \hline
\end{tabular}
}
\end{table}

%% file: figure/rq4n5/rq4_combine.tex
\begin{figure*}[t!]
\centering 
\begin{subfigure}{ 0.25\linewidth}
\centering
\includegraphics[width=\textwidth]{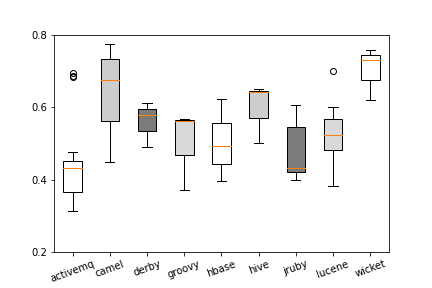}
\caption{\small {\hitrate} of \emph{LIME}}
\end{subfigure}\hfill
\begin{subfigure}{ 0.25\linewidth}
\centering
\includegraphics[width=\textwidth]{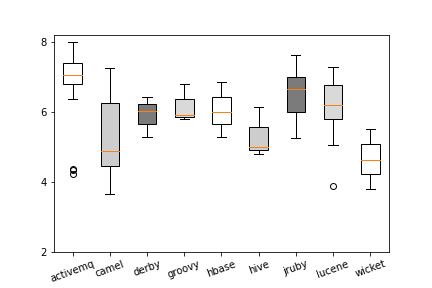}
\caption{\small {\rd} of \emph{LIME}}
\end{subfigure}\hfill
\begin{subfigure}{ 0.25\linewidth}
\centering
\includegraphics[width=\textwidth]{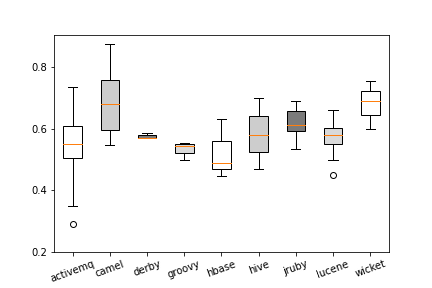}
\caption{\small{ {\hitrate}} of \emph{BreakDown}}
\end{subfigure}\hfill
\begin{subfigure}{ 0.25\linewidth}
\centering
\includegraphics[width=\textwidth]{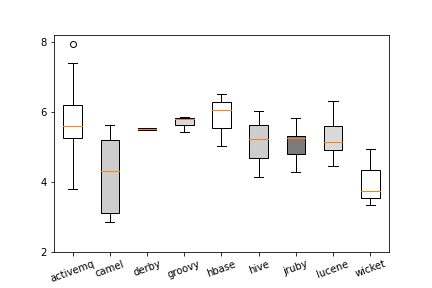}
\caption{\small{ {\rd}} of \emph{BreakDown}}
\end{subfigure}\hfill

\vspace{-0.1in}
\caption{The detailed distributions of {\hitrate} and {\rd} of \emph{LIME} and \emph{BreakDown} on each project under the cross-version defect prediction scenario.}
\label{fig:rq4_comb}
\end{figure*}
\vspace{-0.1in}

%% file: figure/rq4n5/rq5_combine.tex
\begin{figure*}[t!]

\centering 
\begin{subfigure}{ 0.25\linewidth}
\centering
\includegraphics[width=\textwidth]{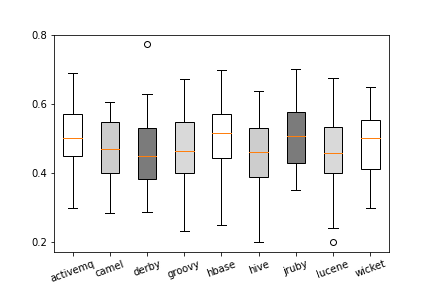}
\caption{\small {\hitrate} of \emph{LIME}}
\end{subfigure}\hfill
\begin{subfigure}{ 0.25\linewidth}
\centering
\includegraphics[width=\textwidth]{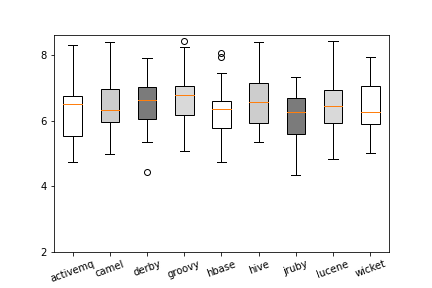}
\caption{\small {\rd} of \emph{LIME}}
\end{subfigure}\hfill
\begin{subfigure}{ 0.25\linewidth}
\centering
\includegraphics[width=\textwidth]{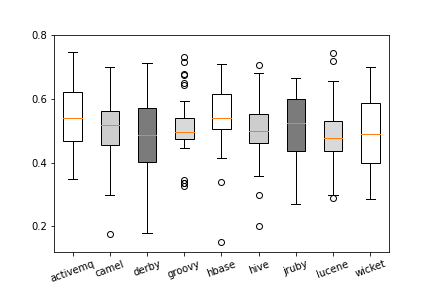}
\caption{\small{ {\hitrate}} of \emph{BreakDown}}
\end{subfigure}\hfill
\begin{subfigure}{ 0.25\linewidth}
\centering
\includegraphics[width=\textwidth]{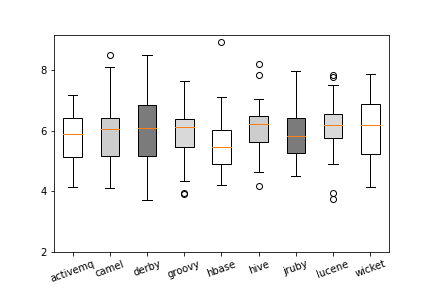}
\caption{\small{ {\rd}} of \emph{BreakDown}}
\end{subfigure}\hfill

\caption{
The detailed distributions of {\hitrate} and {\rd} of \emph{LIME} and \emph{BreakDown} on each project under the cross-project defect prediction scenario.}
\label{fig:rq5_comb}
\end{figure*}
\vspace{-0.1in}

%% file: table/RQ34.tex
\begin{table}[t!]
\centering
\caption{
Average {\hitrate} and {\rd} of the explanations generated by \emph{LIME} and \emph{BreakDown} under different defect prediction scenarios.}
\label{tab:rq34}
\setlength\tabcolsep{3.0pt} 
\begin{tabular}{|c|c|c|c|c|}
\hline
\multirow{2}{*}{\textbf{\begin{tabular}[c]{@{}c@{}}Prediction \\ Scenario\end{tabular}}} & \multicolumn{2}{c|}{\textbf{LIME}} & \multicolumn{2}{c|}{\textbf{BreakDown}}  \\ \cline{2-5} 
  & {\hitrate} & {\rd} & {\hitrate} & {\rd} \\ \hline
\textbf{Cross-Version} & 0.518 & 6.172 & 0.591 & 5.213\\ \hline \hline
\textbf{Cross-Project} & 0.480 & 6.410 & 0.510 & 5.920\\ \hline
\end{tabular}
\end{table}

%% file: figure/diff_n/diff_n.tex
\begin{figure*}[t!]
\centering 
\begin{subfigure}{ 0.25\linewidth}
\centering
\includegraphics[width=\textwidth]{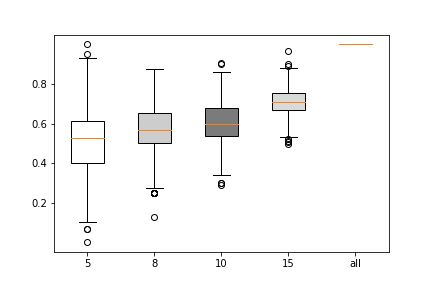}
\caption{\small {\hitrate} of \emph{LIME}}
\end{subfigure}\hfill
\begin{subfigure}{ 0.25\linewidth}
\centering
\includegraphics[width=\textwidth]{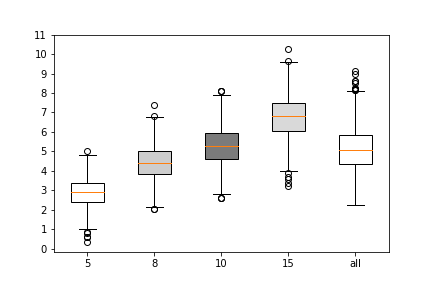}
\caption{\small {\rd} of \emph{LIME}}
\end{subfigure}\hfill
\begin{subfigure}{ 0.25\linewidth}
\centering
\includegraphics[width=\textwidth]{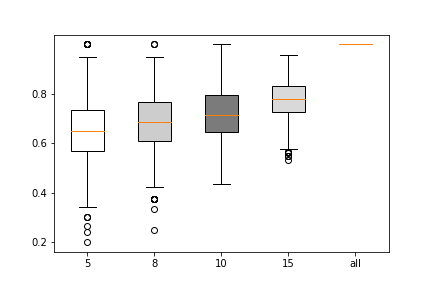}
\caption{\small{ {\hitrate}} of \emph{BreakDown}}
\end{subfigure}\hfill
\begin{subfigure}{ 0.25\linewidth}
\centering
\includegraphics[width=\textwidth]{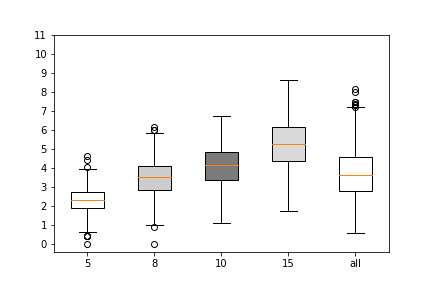}
\caption{\small{ {\rd}} of \emph{BreakDown}}
\end{subfigure}\hfill

\vspace{-0.1in}
\caption{
The distributions of {\hitrate} and {\rd} of \emph{LIME} and \emph{BreakDown} with different top-N features.}
\label{fig:diff_n}
\end{figure*}

%% file: table/discussion2.tex
\begin{table}[t!]
\centering
\vspace{-0.1in}
\caption{
Average {\hitrate} and {\rd} of \emph{LIME} and \emph{BreakDown} with different $N$ features selected.}
\label{tab:diff_ks}
\setlength\tabcolsep{6pt} 
\begin{tabular}{|c|c|c|c|c|}
\hline
\multicolumn{1}{|c|}{\multirow{2}{*}{\textbf{N}}} & \multicolumn{2}{c|}{\textbf{LIME}} & \multicolumn{2}{c|}{\textbf{BreakDown}}  \\ \cline{2-5} 
\multicolumn{1}{|l|}{} & {\hitrate} & {\rd} & {\hitrate} & {\rd}\\ \hline
\textbf{5} & 0.522 & 2.914 & 0.650 & 2.279 \\ \hline
\textbf{8} & 0.571 & 4.416 & 0.686 & 3.446 \\ \hline
\textbf{10} & 0.606 & 5.277 & 0.713 & 4.096 \\ \hline
\textbf{15} & 0.713 & 6.788 & 0.776 & 5.229 \\ \hline
\end{tabular}
\end{table}

%% file: sec/discussion.tex
\section{Discussion}
\label{sec:discussion}

\subsection{Impact of Top-N Features Used in LIME and BreakDown}
\label{sec:dis_issue1}
Following Jiarpakdee et al.~\cite{jiarpakdee2020empirical}, in our experiments, we use top 10 features to generate explanations. However, as both the {\hitrate} and {\rankdiff} can be affected by the number of features used (i.e., $N$), we further investigate whether our findings of \emph{LIME} and \emph{BreakDown} hold when different numbers of features are used. 

For our analysis, we take our RQ1 (Section~\ref{sec:answer_rq2}) as an example to show the impact of different N on the performance of \emph{LIME} and \emph{BreakDown}. 
We follow the same process as described in Section~\ref{sec:answer_rq2} to conduct the experiments with different values of $N$. We experiment $N$ with four values, i.e., 5, 8, 10, and 15. 
For each project, we combine each of the six examined machine learning classifiers (Section~\ref{sec:classifiers}) with each of the five data sampling techniques (Section~\ref{sec:sampling}), in total, there are 30 runs, we use the two tools to generate the explanations for models before and after applying a data sampling technique and further calculate the {\hitrate} and {\rankdiff} values, finally, we average all the {\hitrate} and {\rankdiff} values on each project for each of the two examined model agnostic tools. 
Table~\ref{tab:diff_ks} shows the average {\hitrate} and {\rankdiff} of \emph{LIME} and \emph{BreakDown} with different numbers of top-N features. 
Figure~\ref{fig:diff_n} shows the detailed distribution of {\hitrate} and {\rankdiff} under a different number of top-N features.

As shown in Table~\ref{tab:diff_ks}, overall, with the increase of $N$, both the {\hitrate} and {\rd} increase, this is natural because increasing $N$ will enlarge the search space of \emph{LIME} and \emph{BreakDown}, thus more matches will occur.  
We can see that when $N$ equals to 5, 8, 10, and 15, around 48\%, 43\%, 39\%, and 29\% features in the generated explanations of \emph{LIME} and 35\%, 31\%, 29\%, and 22\% of \emph{BreakDown} are different.  
We have also revisited other three RQs (RQ2-RQ4) by using \emph{LIME} and \emph{BreakDown} with different $N$ and we observe similar results, which indicates \emph{LIME} and \emph{BreakDown} always generates inconsistent explanations regardless of the setting of $N$.


\subsection{Consistency Between the Explanations and the Root Causes of Predicted Bugs}
\label{sec:dis_issue2}

Jiarpakdee et al.~\cite{jiarpakdee2020empirical} used a set of ranked features as the explanation for a prediction. Their human-involved case study showed that 65\% of the participants agree that model-agnostic techniques can generate the time-contrast explanation to answer the why-questions like \textit{Why was file A not classified as defective
in version 1.2 but was subsequently classified as defective in version 1.3?}. 
However, it's still unknown whether the explanations are consistent with the root causes of the buggy instances. To do so, we use the LR-based defect prediction model from Jiarpakdee et al.~\cite{jiarpakdee2020empirical} to predict buggy files on the latest version of each project, we then randomly select 10 correctly predicted buggy instances from each project, in total 90 instances and their explanations generated by both \emph{LIME} and \emph{Breakdown} are collected. For each instance, we trace the data labelling process used in Jiarpakdee et al.~\cite{jiarpakdee2020empirical} to find the linked bug report(s) and we use both a report's content and its corresponding patch(es) to summarize its root cause. 
Finally, we manually (independently by the authors) check whether the generated explanations from \emph{LIME} and \emph{Breakdown} are consistent. 

Our manual analysis shows that none of the explanations generated by these tools can reflect the ground truth root causes of the predicted buggy instances. 
For example, the file ``\textit{ActiveMQConnection.java}'' showed in Figure~\ref{fig:motiv_ex} was labelled as buggy because of two bug reports, i.e., \textit{AMQ-1758}\footnote{https://issues.apache.org/jira/browse/AMQ-1758} and \textit{AMQ-1646}\footnote{https://issues.apache.org/jira/browse/AMQ-1646}, which are caused by an incorrect variable usage and an incorrect condition respectively. 
However, as shown in Figure~\ref{fig:motiv_ex}, the generated explanations are the ordered features and their numerical values, which are unrelated to the root causes, e.g., logic errors, missing API usages, syntax errors, functional errors, etc. 

This result is natural and expected because features used in Jiarpakdee et al.~\cite{jiarpakdee2020empirical} are all high-level software metrics, which can only capture the overall statistical characteristics of software programs. Our analysis confirms with these features, both \emph{LIME} and \emph{Breakdown} cannot generate explanations that can reflect the root causes of buggy instances, which is unreliable to be used in practice. 


\subsection{Threats to Validity}
\label{sec:threats}
\noindent \textbf{Internal Validity.} 
The main internal threat of our study is the limited number of model-agnostic techniques (i.e. \emph{LIME} and \emph{BreakDown}) that we explored. Due to this limitation, we can't generalize our results to all model-agnostic techniques in the file-level defect prediction discipline. 
However, in our future studies, we will explore more techniques and compare the results to \emph{LIME} and \emph{BreakDown}. Furthermore, in this paper, we described a detailed methodology and setup of the experiment and the data set used, allowing other researchers to contribute to our study or further explore the other unexplored techniques.

\noindent \textbf{External Validity.} 
Even though the data sets used in this work are well labelled based on ground truths, the number of the data sets is limited and makes it hard to generalize our results on other data sets and domains. Future work needs to further investigate the study on other data sets. Besides, all the experiment projects are Java projects, although they are popular projects and widely used in existing software bug prediction studies, our findings may not be generalizable to commercial projects.


\noindent \textbf{Construct Validity.} 
To measure the consistency of explanations generated by the same model agnostic technique (i.e., \emph{LIME} and \emph{BreakDown}) under different defect prediction settings, 
we use top 10 features in the explanations to calculate metrics {\hitrate} and {\rankdiff} following Jiarpakdee et al.~\cite{jiarpakdee2020empirical}. 
With different number of features used, the {\hitrate} and {\rankdiff} of two explanations can be different, which could affect our findings. However, as we shown in Section~\ref{sec:dis_issue1}, \emph{LIME} and \emph{BreakDown} ways generates inconsistent explanations regardless of the number of features used.


%% file: sec/related.tex
\section{Related Work on Explainable Defect Prediction}
\label{sec:related}
As analytical modelling advances in the software engineering domain, the lack of explainability of analytical models becomes more problematic. Recent studies show the importance and need for such explanations~\cite{radford2017learning}. Even more, as Dam et al., Lewis et al., Menzies and Zimmermann emphasize, these analytical model explanations need to be actionable to provide the most value and practical use to both practitioners and software engineers~\cite{radford2017learning, lewis2013does, 6547619,JiarpakdeeTG21}. 

Many efforts have been done to build explainable software defect prediction models~\cite{humphreys2019explainable,jiarpakdee2020empirical,khanan2020jitbot,pornprasit2021jitline,9193975,lundberg2017unified,ribeiro2018anchors}. 
Jiarpakdee et al.~\cite{JiarpakdeeTG21} conducted a qualitative survey that investigates developers' perception of defect prediction goals and their explanations. 
The results of their experiments showed that majority of the respondents believed that software defect prediction is very important and useful and  \emph{LIME} and 
\emph{BreakDown} are ranked as the top two approaches among a list of explanation generation approaches, in terms of the usefulness, quality, and insightfulness of the explanations. 
Humphreys and Dam~\cite{humphreys2019explainable} proposed an explainable deep learning defect prediction model which exploits self-attention transformer encoders. By using self-attention transformer encoders, the model can disentangle long-distance dependencies and benefit from its regularizing effect.
Also, they can normalize correlations that are inversely proportional to the prediction for more useful data. Jiarpakdee et al. \cite{jiarpakdee2020empirical} used \emph{LIME} and \emph{BreakDown} to generate explanations on file-level defect prediction models that show which metrics are associated with buggy predictions. 
Khanan et al. \cite{khanan2020jitbot} proposed an explainable JIT-DP framework, \emph{JITBot}, that automatically generates feedback for developers by providing risks, and explaining the mitigation plan of each commit.
They used Random forest classifier for risk introducing commit prediction and leveraged model-agnostic technique, i.e. \emph{LIME}, to explain the prediction results. 
Pornprasit and Tantithamthavorn~\cite{pornprasit2021jitline} proposed \emph{JITLine}, which ranks defective lines in a commit for finer granularity. 
With \emph{JITLine}, they are able to predict both defect-introducing commits and identify lines that are associated with the commit. They exploit Bag-of-Token features extracted from repositories and apply them on machine learning classifiers to calculate the defect density of each commit. Then, they use defect density scores to rank different lines of the commit as risky. 
Wattanakriengkrai et al. \cite{9193975} proposed a framework called \emph{LINE-DP}, which applies \emph{LIME} on a file-level prediction model trained with code token features.
The explanation generated from \emph{LIME} will show which code tokens are introducing bugs in the file.
Then they use these explanations to identify a line buggy if the line contains bug-prone tokens. 
Lundber and Lee \cite{lundberg2017unified} proposed {SHAP} which is a model-agnostic technique that works similarly to \emph{BreakDown}, however instead of using the greedy strategy, it uses game theory to calculate the contribution probability of each feature to the final prediction of the prediction model. 
Ribeiro et al. \cite{ribeiro2018anchors} proposed Anchors. \emph{Anchor} is an extension of \emph{LIME} generating rule-based explanations using decision rules. These rules are if-then rules (anchors) that have high confidence (at least 95\% confidence and highest coverage if more than one rule has the same confidence value). In other words, only the selected features by \emph{anchor} affect the final prediction. 

Recently, Reem~\cite{aleithan2021explainable} conducted the first study to manually check whether the explanations generated by \emph{LIME} and \emph{BreakDown} are the same as the root cause of the bugs for change-level defect prediction models. Their results showed that both \emph{LIME} and \emph{BreakDown} fail to explain the root causes of predicted buggy changes. 
In this work, we conduct an empirical study to analyze the reliability and stability of model-agnostic based explanation generation techniques, i.e., \emph{LIME} and \emph{BreakDown} on software defect prediction under various settings at file-level defect prediction and we have also conducted the same manual analysis as Reem~\cite{aleithan2021explainable} on file-level defect prediction models and our manual analysis confirms the same finding, i.e., both \emph{LIME} and \emph{BreakDown} fail to explain the root causes of predicted buggy instances.

%% file: sec/conclusion.tex
\section{Conclusion}
\label{sec:conclusion}
In this paper, we investigate the reliability and stability of model-agnostic based explanation generation techniques, i.e., \emph{LIME} and \emph{BreakDown}, under different software defect prediction settings. 
Our experiments on 32 versions of defect prediction data from nine open-source projects show that neither \emph{LIME} nor \emph{BreakDown} can generate consistent explanations under different defect prediction settings, thus both are unreliable to be used in practice. 
{In addition, our manual analysis confirms that none of the generated explanations can reflect the root causes of the predicted defects.} 
Thus, {contrary to the claim of Jiarpakdee et al.~\cite{jiarpakdee2020empirical}, our study suggests 
that model-agnostic techniques
are neither reliable nor stable to be used for explanation generation for defect prediction.} 

In the future, we plan to examine the reliability and stability of model-agnostic techniques used in other software engineering tasks and explore more reliable explanation generation techniques for prediction tasks in the software engineering domain.
